\documentclass{rmf-d}
\usepackage{nopageno,rmfbib,multicol,times,epsf,amsmath,amssymb,cite}
\usepackage[T1]{fontenc} %Especial para español (for spanish)
\usepackage[]{caption2}
\usepackage{graphicx}
\usepackage{blindtext}
\usepackage{color}
\usepackage{booktabs}
\usepackage{hyperref}
\usepackage{multirow}

\def\rmfcornisa{Gravitation, Mathematical Physics and Field Theory \hfill\rmf\ {\bf ??} (*?*) ???--???
\hfill APR, 2025}

\def\rmfcintilla{{\it Rev.\ Mex.\ Fis.\/} {\bf ??} (*?*) (????) ???--???}
\clearpage 
\begin{document}
\markboth{  José Blanco, Cuauhtémoc Campuzano, J. C. Corona-Oran and Víctor H. Cárdenas}{Perturbation of a cosmological Stealth field.}

%%%%%
%
% Please provide the following information
%
%%%%%
\title{   Non-geometrical perturbation on homogeneous stealth dust
\vspace{-6pt}}
\author{  José Blanco$^{1,3}$, Cuauhtemoc Campuzano$^{1}$ J. C. Corona-Oran $^{2}$ and Víctor H. Cárdenas$^3$ }
\address{$^1$ Universidad Veracruzana, Facultad de Física, 91223 Xalapa-Enríquez, Ver., México  }
\address{ $^2$Universidad Autónoma del Estado de México, Facultad de Ciencias, 50200 Toluca de Lerdo, Méx., México}
\address{ $^3$ Universidad de Valparaíso, Instituto de Física y Astronomía, Av. Gran Bretana 1111, Valparaíso, Chile}

%%%%%
%
% Use as many authors and addresses as required
%
%%%%%

\maketitle
%%%%%
%
% To be filled by the Editorial Team of RMF, RMF-E 
% and SRMF
%
%%%%%
\recibido{16 DEC 2024}{17 APR 2025
\vspace{-12pt}}
\begin{abstract}
\vspace{1em} 
%%%%%
%
% Provide your abstract
%
%%%%%

In this work, we study a non-geometrical perturbation to the stealth field, which means the background remains invariant. The stealh is homogeneous in a universe whose source is dust and demand that perturbation unchanged density. As a regular procedure, we introduce a parameter $\lambda$ to perturb the scalar field equation and get an intriguing expression of the equation, similar to a series expansion in $\lambda$. From this procedure, we distinguish and approach to discriminate solutions, and the numerical solutions show that the most significant contribution to the solution comes from the linear term of $\lambda$.

\vspace{1em}
\end{abstract}
\keys{ \bf{\textit{scalar fields, cosmology, numerical relativity.
}} \vspace{-8pt}}
\begin{multicols}{2}

\section{Introduction}

Currently, the most striking discovery is the accelerated expansion of the universe, which earned the 2011 Nobel Prize in Physics \cite{riess1998observational,perlmutter1999measurements,koyama2016cosmological}, and explaining it is possibly the main challenge of theoretical physics. This phenomenon and discrepancies between theories and observations that agree with General Relativity (GR) lead us to question the validity of the standard cosmological model and to explore alternative theories to GR \cite{joyce2015beyond}. As far as we know, success of GR theory in explaining gravity is based on observations made in the vicinity of the solar system (i.e., under weak gravity conditions), as well as laboratory experiments (at millimeter scale) and tests of gravitational wave emission by binary pulsars \cite{will2014confrontation,schutz2022first}. Even more in context, the cosmological models coming from GR to explain the acceleration have theoretical limitations.
\\

The best fit of the observational data is given by a cosmological model known as $\Lambda$-Cold Dark Matter ($\Lambda$CDM), which comes from GR. In such a model, the cosmological constant ($\Lambda$) is introduced into Einstein’s equations to explain the acceleration. It is assumed to be the origin of dark energy (DE) that exerts negative pressure. Despite this, it is spectacularly incomplete and unsatisfactory from a theoretical standpoint due to the lack of physical justification. Furthermore, the model faces two of the most successful theories today: GR and field theory. Relativity predicts a smaller value of the cosmological constant than compared to what high-energy physics suggests, resulting in a discrepancy of more than 120 orders of magnitude between both predictions. It is also important to note that differences are found between local measurements of the Hubble constant and statistical inferences obtained from the cosmic microwave background and observations of distant supernovae \cite{di2021realm}. If the existence of dark energy is confirmed, our limited knowledge of its nature is a cause for concern \cite{peebles2003cosmological}.
\\

Regardless of the above, the greatest challenge is to find a theory that explains the nature of the different components of $\Lambda$CDM, namely dark energy, dark matter, and the inflation field (necessary for structure formation and describing a phase of extremely highly acceleration in the early universe). In this context, the standard model of cosmology arises from a strong and questionable extrapolation of our limited knowledge of gravity, as GR has not been tested at galactic and cosmological scales. 
\\

Among the alternatives, there are models based on Modified Gravity theories (MOG) that offer different options to explain the origin of the cosmological constant and the late acceleration of the universe. However, by significantly modifying gravity on cosmological scales, they face problems in meeting the strict constraints of the solar system. In this regard, there are different proposals to avoid modifications of gravity on small scales \cite{joyce2015beyond}, among which massive gravity, bi-gravity, and degenerate higher-order scalar-tensor theories (DHOST) stand out \cite{koyama2016cosmological, sotiriou2010f}.
\\

Models based on MOG adequately describe gravitational singularities and propose explanations to understand the dark sector of the universe \cite{delaunay2021stealth, campuzano2016mimicking}. However, in the study of MOG, a significant problem arises in finding solutions to the associated partial differential equations, as coupling a field introduces more degrees of freedom. The stealth solutions studied belong to MOG theories; they are configurations interpreted as non-gravitating matter and do not alter the background spacetime \cite{ayon2005nonminimally, ayon2013conformal, alvarez2019cosmological}.
\\

One of the widely used formalisms in physics to extract relevant information from a phenomenon is to perturb the system under study. After perturbing the stealth, there is no reason the field remains in that configuration, so if any way to detect scalar fields exists, it would be possible to detect these fields gravitationally by perturbing them. Our purpose is to find an approximate solution for the stealth by modifying its exact solution using the simplest perturbation $\phi \rightarrow \phi + \lambda \delta\phi$, where $\phi$ is the unperturbed function, $\lambda$ is a parameter related to the intensity of the alteration to the system $0<\lambda\leq 1$.
\\

In Section 2, we provide a brief general review of the action used to obtain the equations that describe the homogeneous and isotropic universe we study, minimally coupled to gravity and with sources. Section 3 obtains the field equations when gravity is non-minimally coupled to the scalar field. We use the Friedmann-Lemaître-Robertson-Walker (FLRW) metric, a homogeneous scalar field, and the dust tensor as a source. In Section 4, we perturb the scalar field in the equations, obtain its numerical solutions, and use them to construct the associated potentials. Finally, in Section 5, we present the discussion and conclusions.

\section{Stealth Fields}
The general framework for addressing the problem, which involves a gravitational field with sources non-minimally coupled to a scalar field, is described by the action,
\begin{equation}
\begin{aligned}
S[g_{\mu\nu},\phi] = \int_{M} d^4 x \sqrt{-g} \bigg( \frac{1}{2\kappa} R + \mathcal{L}_m
- \frac{1}{2} \zeta R \phi^2 \\
- \frac{1}{2} \partial_{\mu} \phi \, \partial^{\mu} \phi - V(\phi)  \bigg),
\end{aligned}
\label{1}
\tag{1}
\end{equation}
above, the first expression inside the parentheses is the Einstein-Hilbert term, where \( R \) is the Ricci scalar, \( \kappa = \frac{8\pi G}{c^4} \), \( G \) is the gravitational constant, \( c \) is the speed of light, the second term is the matter Lagrangian, followed by the non-minimal coupling; where \( \zeta \) the coupling constant (\( \zeta \neq 0 \) because if  \( \zeta = 0 \) we have minimal coupling), and the rest belongs to the scalar field \( \phi \), including the self-interaction potential \( V(\phi) \). 

From the variation of the action with respect to the metric, we obtain the following set of equations:
\begin{equation}
G_{\mu\nu}-kT_{\mu\nu}^{(m)}=\kappa T_{\mu\nu}^{(S)}, 
\label{2}
\tag{2}
\end{equation}
where, \( G_{\mu\nu} \) is th Einstein tensor, \( T_{\mu\nu}^{(m)} \) is the energy-momentum tensor of matter, and \( T_{\mu\nu}^{(S)} \) is the energy-momentum tensor of a scalar field (including the non-minimal coupling), explicitly,

\begin{equation}
\begin{aligned}
T_{\mu\nu}^{(S)} = \nabla_{\mu} \phi \nabla_{\nu} \phi 
- \left[ \frac{1}{2} \nabla_{\lambda} \phi \nabla^{\lambda} \phi + V(\phi) \right] g_{\mu\nu} \\
+ \zeta \left( G_{\mu\nu} + g_{\mu\nu} \Box - \nabla_{\mu} \nabla_{\nu} \right) \phi^2.
\end{aligned}
\label{3}
\tag{3}
\end{equation}

A scalar field is stealth if is non trivial and satisfies the following equation:
\begin{equation}
T_{\mu\nu}^{(S)}=0, 
\label{4}
\tag{4}
\end{equation}
the solutions to the above equations are configurations of matter that do not gravitate, known as stealths. On the other hand, from the variation of the scalar field, we obtain the equations that describe its dynamics: 
\begin{equation}
\Box \phi = \zeta R \phi + \frac{dV(\phi)}{d\phi}. 
\label{5}
\tag{5}
\end{equation}
It is worth mentioning that, due to the invariance under diffeomorphisms of the action \eqref{1}, the solutions of the equations \eqref{3} necessarily imply that the equations for the dynamics of the scalar field \eqref{5} are satisfied \cite{alvarez2019cosmological}.

\section{Stealth Cosmological Field}

In this section, we briefly review the stealth fields to be studied, namely, scalar fields in a homogeneous and isotropic universe, described by the FLRW metric:

\begin{equation}
ds^2=-dt^2+\frac{dr^2}{1-k r^2} + a^2\left( r^2 d\Theta^2 +r^2 \sin^2(\Theta) d\phi^2 \right)
\label{6}
\tag{6}
\end{equation}
\( a=a(t) \) is the scale factor (the value \( a(t)=0 \) corresponds to the moment of the Big Bang, which implies a singularity. In contrast, \( a(t)=1 \) represents the value of the scale factor for the current universe) and \( k \) describes the curvature of spacetime, whose values can be \( k=-1, 0, \) or \( 1 \). In this work, we will only consider the case of a flat universe, that is, \( k=0 \). The source is a perfect fluid with an energy-momentum tensor:

\begin{equation}
T_{\mu\nu}^{(m)}=(\rho+p) u_{\mu} u_{\nu}+p g_{\mu\nu},
\label{7}
\tag{7}
\end{equation}
where \( \rho=\rho(t) \) is the energy density, \( p=p(t) \) is the intrinsic pressure of the universe, and \( u^{\mu} \) denotes the 4-velocities of the fluid, with the property that \( u^{\mu} u_{\nu}=-1 \).
\\

 Starting from equation (\ref{3}), and assuming the validity of Einstein's equations together with the condition \eqref{4}, we obtain a system of equations from which the expressions for \( \rho \) and \( p \) can be derived.

\begin{equation}
\zeta k \rho = -\frac{d}{dt} \ln(\phi) \cdot \frac{d}{dt} \ln\left(a^{6 \zeta} \phi^{1/2}\right) - \frac{V(\phi)}{\phi^2}
\label{8}
\tag{8}
\end{equation}
and
\begin{equation}
\zeta k p = -\zeta \frac{d}{dt} \ln(\phi) \cdot \frac{d}{dt} \ln\left(\frac{\phi^{(1-4\zeta)/2\zeta}}{a^4 \dot{\phi^2}}\right) + \frac{V(\phi)}{\phi^2}.
\label{9}
\tag{9}
\end{equation}
We are interested in the case when the pressure is zero (\( p=0 \)), known as dust. The system is reduced by isolating the term \( \frac{V(\phi)}{\phi^2} \) in equation \eqref{9} and substituting it into equation \eqref{8}. Now, we reduced to solve the following equation 
\begin{equation}
k \rho(t) = \left(\frac{d}{dt} \ln(\phi)\right) \cdot \frac{d}{dt} \ln \left(\frac{\dot{\phi} \phi^{(2\zeta-1)/\zeta}}{a^2}\right)^2.
\label{10}
\tag{10}
\end{equation}

 For convenience, we make the change of variable \( \phi(t) \rightarrow \phi(a(t)) \), due to this change, all the variables \( \phi, \rho, p, V \) now depend on the scale factor. Additionally, the energy density for the dust case is given by \( \rho = \rho_0 a^{-3} \). From here, we can rewrite equation \eqref{10} as
%
%In our case, the scale factor is $a(t) = (a_1 t + a_0)^{2/3}$ where \( a_1^2 = \frac{\kappa \rho_0}{3} \). 
%
\begin{equation}
\frac{\phi''}{\phi} - \frac{3}{2a}  \left( \frac{\phi'}{\phi} \right) + \frac{(2\zeta-1)}{2\zeta} \left( \frac{\phi'}{\phi} \right)^2 = \frac{3}{2a^2},
\label{12}
\tag{12}
\end{equation}
where \( \frac{d\phi}{da} = \phi' \). In the described scenario, there are four branches of solutions that depend on the range of \( \zeta \) reported in \cite{alvarez2019cosmological} and shown in Table I.

\section{Non-geometrical Stealth Field Perturbations}

Perturbation is a mathematical method used to find approximate solutions to complicated equations. It assumes knowledge of the exact solution to a problem and studies whether a small modification yields an approximate solution or a solution to a new problem that is a slight modification of the known one. This method is challenging to implement for GR because one must also make the perturbation on the background space, which changes the point at which a tensorial quantity is evaluated (a good discussion of the perturbation theory in GR can be found in the reference \cite{unanue2011revision}). However, in this work, a condition we require is that spacetime remains invariant during the perturbation, which allows us to implement the classical perturbation method to obtain the quantity \(F\) from the exact solution \(F_0\), expressing it as a power series of the parameter \(\lambda\), i.e., $F = F_0 + \lambda F_1 + \lambda^2 F_2 + O(\lambda^3)$.
\\

As mentioned, we study a perturbed stealth solution by modifying the exact one without altering the background space. The way to achieve this is by replacing \(\phi \to \phi + \lambda \delta\phi\) in equation \eqref{12}, yielding 

\begin{equation}
\begin{aligned}
\frac{\phi'' + \lambda \delta\phi''}{\phi + \lambda \delta\phi} 
&- \frac{3}{2a} 
\frac{\phi' + \lambda \delta\phi'}{\phi + \lambda \delta\phi} \\
&+ \frac{(2\zeta - 1)}{2\zeta} 
\left( 
\frac{\phi' + \lambda \delta\phi'}{\phi + \lambda \delta\phi} 
\right)^2 
= \frac{3}{2a^2},
\end{aligned}
\label{13}
\tag{13}
\end{equation}
expanding the above equation we get
\begin{equation}
\begin{aligned}
&\phi\phi''-\frac{3}{2a}\phi\phi'+\frac{2\zeta-1}{2\zeta}\phi'^2+\\
&+\lambda \big[\delta\phi\phi''+\phi\delta\phi''-\frac{3}{2a}\big(
\delta\phi\phi'+\phi\delta\phi'\big)+\frac{2\zeta-1}{\zeta}\delta\phi'\phi' \big]\\
&+\lambda^2 \big[\delta\phi\delta\phi''-\frac{3}{2a}\delta\phi\delta\phi'+\frac{2\zeta-1}{2\zeta}\delta\phi'^2\big]\\
&=\frac{3}{2a^2}(\phi^2+2\lambda\phi\delta\phi+\lambda^2\delta\phi^2).
\end{aligned}
\label{14}
\tag{14}
\end{equation}

The current problem is to find the field \(\delta\phi\). In equation \eqref{14}, we impose the condition that matter density does not change; physically, one can achieve this in different ways, similar to perturbing a liquid without increasing its volume. While \(\lambda\) quantifies how large or small the perturbation is, in our case, we focus on the behavior of \(\delta\phi\) rather than its nature.

\subsection{Perturbation Solutions}

The equation (\ref{14}) looks like a series expansion in terms of $\lambda$. The zero-order term is the stealth equation, while the coefficients of linear and quadratic terms are the perturbed equation (this is more obvious when $\lambda =1$). In this case, the introduction of the parameter $\lambda =1$ shows that the terms of the square of l are equal to those of the equation for the stealth. Because of its particular form, we wonder what would happen if we considered just the linear coefficient; instead of taking both, we found that the contribution most important to the solution comes from the linear coefficient. 

The perturbed equation is
\begin{equation}
\begin{aligned}
&\lambda \big[\delta\phi\phi''+\phi\delta\phi''-\frac{3}{2a}\big(
\delta\phi\phi'+\phi\delta\phi'\big)+\frac{2\zeta-1}{\zeta}\delta\phi'\phi' \big]\\
&+\lambda^2 \big[\delta\phi\delta\phi''-\frac{3}{2a}\delta\phi\delta\phi'+\frac{2\zeta-1}{2\zeta}\delta\phi'^2\big]\\
&=\frac{3}{2a^2}(2\lambda\phi\delta\phi+\lambda^2\delta\phi^2),
\end{aligned}
\label{15}
\tag{15}
\end{equation}
which we solve using Runge-Kutta-Fehlberg (RKF45) methods to second-order ordinary differential equations \cite{Hairer2008}. 

After perturbing the scalar field, we will observe $\phi+\delta\phi$, so  it is worth contrasting with $\phi$ ( the following Table provides a summary of solutions reported in \ref{1}). In the following graphs, we show this contrast for all branches with the different parameters

\end{multicols}

\begin{table}[H]
\centering
\small % Reduce font size 
\begin{tabular}{|c|c|c|}
\hline
\textbf{Range of $\zeta$} & \textbf{Solution} & \textbf{Constants and variables involved} \\ \hline
$\zeta \in (-\infty, 0) \cup (12/73, \infty) - 1/4$ (Branch 1) 
& $\phi(a) = \Phi_0 \left( \left(\frac{a}{a_0}\right)^{5/4+\beta} - \left( \frac{a}{a_0} \right)^{5/4-\beta} \right)^{\frac{2\zeta}{4\zeta -1}}$ 
& $a_0 = \left( \frac{\phi_0}{\phi_1} \right)^{1/2\beta}$, \\
& & $\beta = \frac{\sqrt{\zeta(73\zeta - 12)}}{4\zeta}$,\\ & & $\Phi_0 = \left(\phi_0 ^{\frac{5/4 + \beta}{2\beta}} -\phi_1 ^{\frac{5/4- \beta}{2\beta}}\right)^{\frac{2\zeta}{4\zeta -1}}$ \\ \hline

$\zeta = 12/73$ (Branch 2) 
& $\phi(a) = (\phi_1 \ln(a) - \phi_0)^{-24/25} a^{-6/5}$ 
&  \\ \hline

$0 < \zeta < 12/73$ (Branch 3) 
& $\phi(a) = \phi_0  \left( a^{5/2}(\phi_1 \sin(\gamma) - \phi_2 \cos(\gamma )) \right)^{\frac{2\zeta}{4\zeta - 1}}$ 
& $\gamma(a) =\frac{\sqrt{12-73\zeta}}{4\sqrt{3}}ln(a)$  \\ \hline

$\zeta = 1/4$ (Branch 4) 
& $\phi(a)_{\pm} = \phi_1 e^{\pm\phi_0 a^{1/2}} a^{-3/2}$ 
&  \\ \hline
\end{tabular}
\caption{The table presents the solutions from [Álvarez, Campuzano, Cárdenas 2019]. $\phi_0$ and $\phi_1$ are integration constants.}
\label{table:solutions}
\end{table}

\begin{multicols}{2}
    
\subsubsection*{Branch 1}
\begin{figure}[H]
    \centering
    \includegraphics[width=\linewidth]{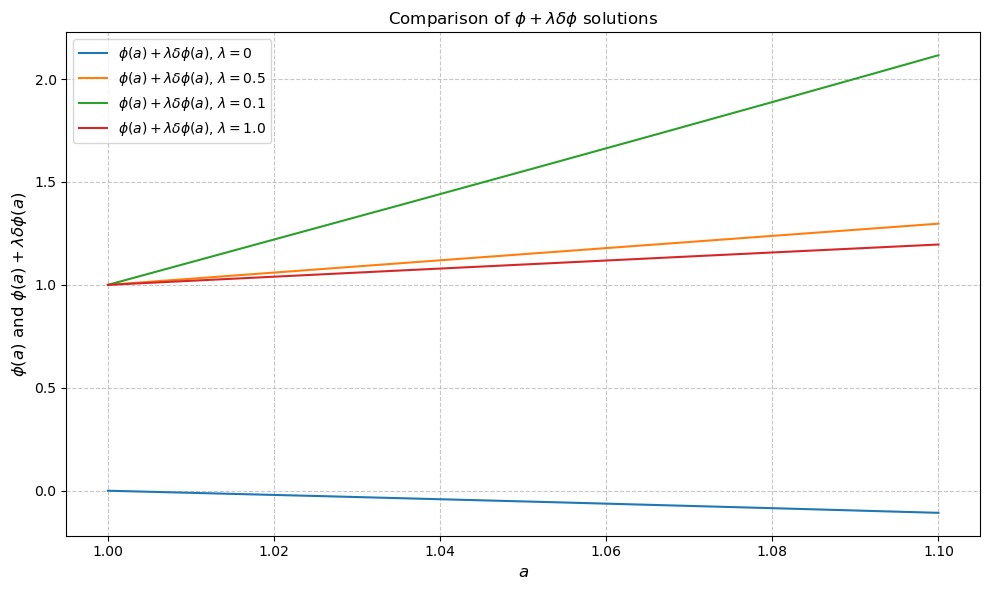}
    \caption{  We show the solution  $\phi + \delta \phi$ with different values of $\lambda$, as a function of the scale factor $a$, over the interval $[1.0, 1.1]$. The initial conditions are $\delta\phi(0.01) = 0$, $\delta\phi'(0.01) = -1$ and the parameters $\zeta = 0.3$, $\phi_0 = 1$ and $\phi_1 = 0.5$.}
    \label{fig:b1}
\end{figure}

\subsubsection*{Branch 2}
\begin{figure}[H]
    \centering
    \includegraphics[width=\linewidth]{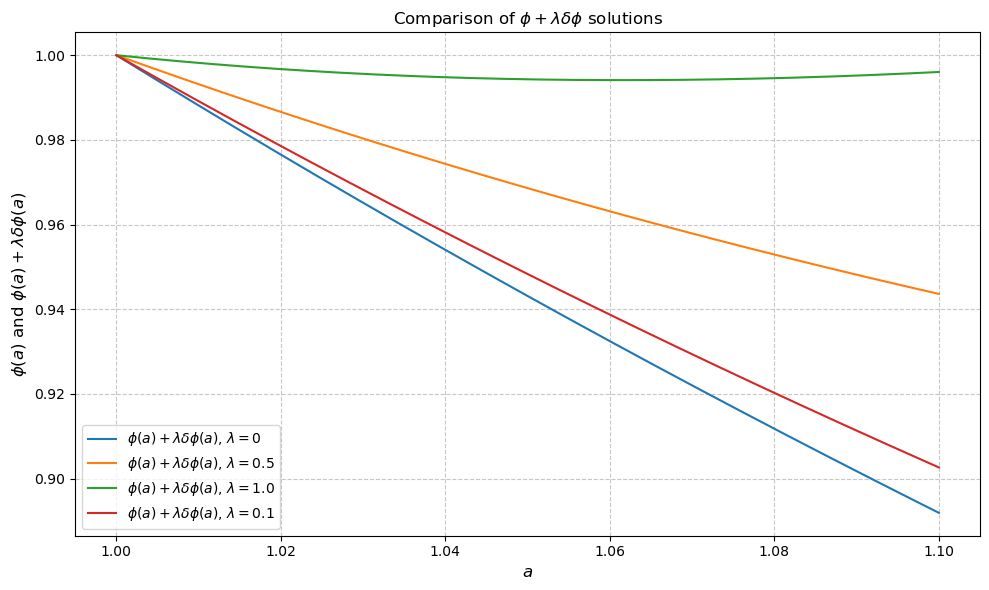}
    \caption{  We show the solution  $\phi + \delta \phi$ with different values of $\lambda$, as a function of the scale factor $a$, over the interval $[1.0, 1.1]$. The initial conditions are $\delta\phi(0.01) = 0$, $\delta\phi'(0.01) = 0$ and the parameters $\zeta = \frac{12}{73}$, $\phi_0 = -1$ and $\phi_1 = -1$.}
    \label{fig:b2}
\end{figure}

\subsubsection*{Branch 3}
\begin{figure}[H]
    \centering
    \includegraphics[width=\linewidth]{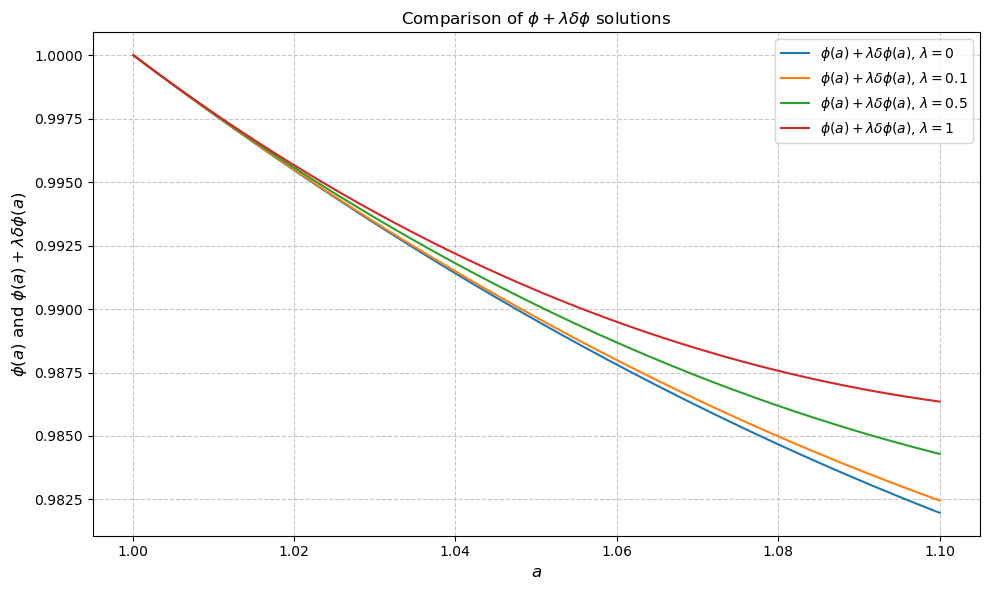}
    \caption{  We show the solution  $\phi + \delta \phi$ with different values of $\lambda$, as a function of the scale factor $a$, over the interval $[1.0, 1.1]$.The initial conditions are \(\delta\phi(0.01) = 1\),  \(\delta\phi'(0.01) = 1\) and the parameters \(\zeta = 0.1\), \(\phi_0 = 1\), \(\phi_1 = -1\) and \(\phi_2 = -1\).}
    \label{fig:b3}
\end{figure}

\subsubsection*{Branch 4}
\begin{figure}[H]
    \centering
    \includegraphics[width=\linewidth]{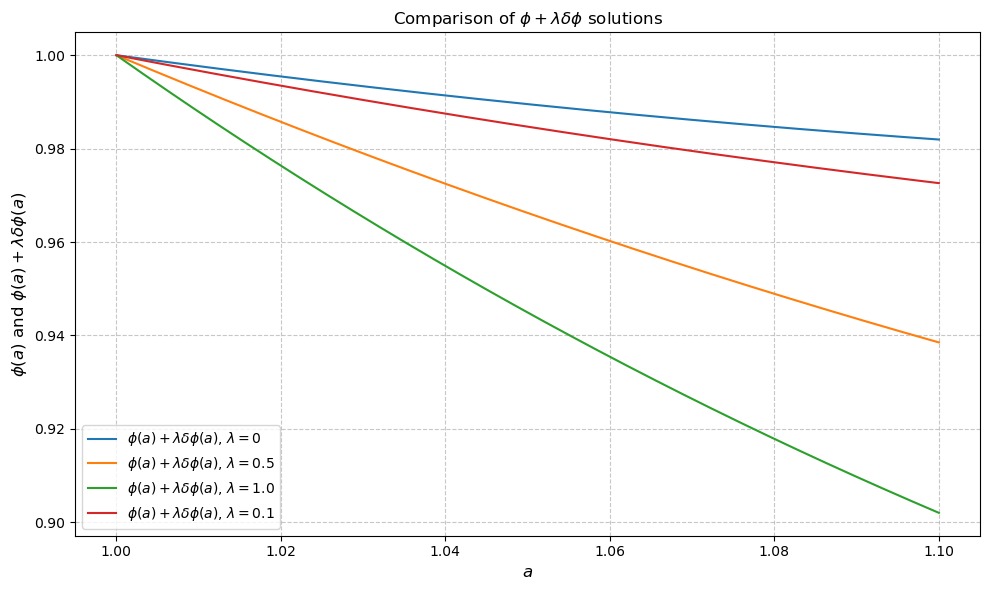}
    \caption{  We show the solution  $\phi + \delta \phi$ with different values of $\lambda$, as a function of the scale factor $a$, over the interval $[1.0, 1.1]$. The initial conditions are \(\delta\phi(0.01) = 1\),  \(\delta\phi'(0.01) = -1\) with \(\zeta = 0.25\), \(\phi_0 = 0\) and \(\phi_1 = -1\) . }
    \label{fig:b4}
\end{figure}

In branches 1 and 2, the numerical solution has unphysical behavior to $\lambda =1$; the solution is closer to the exact than $\lambda < 1$ which means more significant perturbation less changes the system. The solutions behave as expected for branches 3 and 4; small perturbations mean fewer changes to the system.

As we pointed out above, we found it intriguin that the quadratic term in $\lambda$ replays the equation for stealth without perturbing. Because of that, we ask if taking the linear term in (\ref{15}) is enough to consider this a good approximation to the solution.We contrast the solutions between the linear term and all terms to get an answer, shown in the following tables, in which we present three values of the linear and complete solutions (normalized) as a function of $a$, computed with a step size of 0.0000001.

\subsubsection*{Branch 1}

\begin{table}[H]
    \centering
    \begin{tabular}{cccc}
        \toprule
        $\lambda$ & $a$ & $\hat{\phi}(linear)$ & $\hat{\phi}(complete)$ \\
        \midrule
        0.1 & 1.0000001  & 0.9999999760  & 0.9999999760  \\
        0.1 & 1.5 & 0.9895267933  & 0.9896527704  \\
        0.1 & 1.1        & 0.9819492944  & 0.9824370606  \\
        \bottomrule
    \end{tabular}
    \caption{Comparison between different values of \(\hat{\phi}\) (linear) and \(\hat{\phi}\) (complete), where $\hat{\phi}= \phi + \lambda \delta \phi$.}
    \label{tab:datos}
\end{table}

\subsubsection*{Branch 2}

\begin{table}[H]
    \centering
    \begin{tabular}{cccc}
        \toprule
        $\lambda$ & $a$ & $\hat{\phi}(linear)$ & $\hat{\phi}(complete)$ \\
        \midrule
        0.1 & 1.0000001 & 0.9999760063 & 0.9999760063 \\
        0.1 & 1.5 & 0.9895178152 & 0.9896435279 \\
        0.1 & 1.1 & 0.9819492944 & 0.9824360391 \\
        \bottomrule
    \end{tabular}
    \caption{Comparison between different values of \(\hat{\phi}\) (linear) and \(\hat{\phi}\) (complete), where $\hat{\phi}= \phi + \lambda \delta \phi$.}
    \label{tab:phi_lambda}
\end{table}

\subsubsection*{Branch 3}
\begin{table}[H]
    \centering
    \begin{tabular}{cccc}
        \toprule
        $\lambda$ & $a$ & $\hat{\phi}(linear)$ & $\hat{\phi}(complete)$ \\
        \midrule
        0.1 & 1.0000001 & 0.9938291366 & 0.9938291366 \\
        0.1 & 1.5 & 0.9897829985 & 0.9896993123 \\
        0.1 & 1.1 & 0.9889763119 & 0.9878341094 \\
        \bottomrule
    \end{tabular}
    \caption{Comparison between different values of \(\hat{\phi}\) (linear) and \(\hat{\phi}\) (complete), where $\hat{\phi}= \phi + \lambda \delta \phi$.}
\end{table}

\subsubsection*{Branch 4}
\begin{table}[H]
    \centering
    \begin{tabular}{cccc}
        \toprule
        $\lambda$ & $a$ & $\hat{\phi}(linear)$ & $\hat{\phi}(complete)$ \\
        \midrule
        0.1 & 1.0000001 & 0.9988000023 & 0.9988000023 \\
        0.1 & 1.5 & 0.9823917582 & 0.9823974198 \\
        0.1 & 1.1 & 0.9886481641 & 0.9871320458 \\
        \bottomrule
    \end{tabular}
    \caption{Comparison between different values of \(\hat{\phi}\) (linear) and \(\hat{\phi}\) (complete), where $\hat{\phi}= \phi + \lambda \delta \phi$.}
\end{table}

It is evident that with small values, there are no differences between the solutions up to 
$10^{-10}$; then all contributions come from the linear term.

For this work, we get no information from the self-interaction potential; however, we show some typical cases in what follows.

%Now, from Eqs. (\ref{8}) and (\ref{9}) once we get \( \delta\phi \) and \( \delta\phi' \), we show the potential \( V(\phi + \delta\phi) \) for its solution and compare it with the lineal approximation. 
%%Revisar tamaño fuente labels graphs
\subsubsection*{Branch 1}
\begin{figure}[H]
 \includegraphics[width=\linewidth]{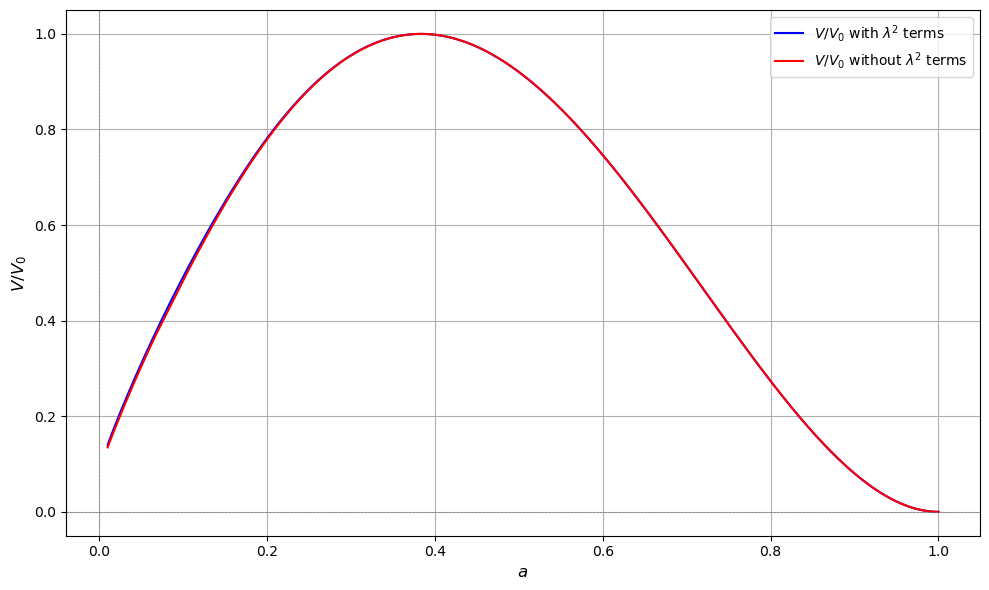}
 \caption{\small{Branch 1. We show the potential \(V(\phi + \delta \phi)\) for the solution and contrast it with the lineal term. In the graph \(V_0\) is the greatest valor of $V$ over the interval $[0.01, 1]$. The initial conditions are $\delta\phi(0.01) = 0$, $\delta\phi'(0.01) = -1$, and with the parameters $\zeta = 0.3$, $\phi_0 = 1$ and $\phi_1 = 0.5$. The blue line is the solution potential and the red is the lineal approximation.}}\label{fig9}
\end{figure}

\subsubsection*{Branch 4}
\begin{figure}[H]
 \includegraphics[width=\linewidth]{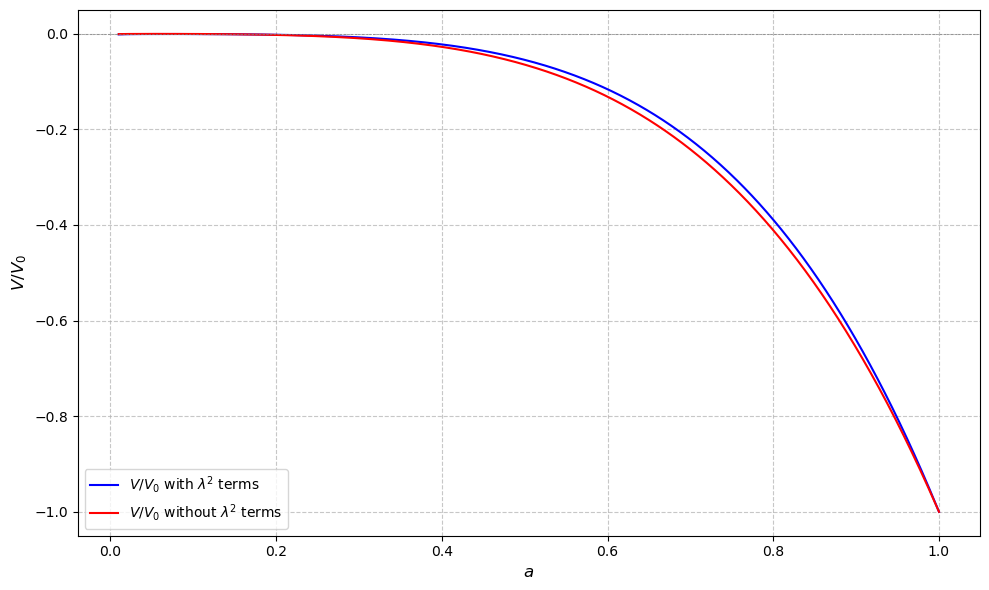}
 \caption{\small{Branch 4.  We show the potential \(V(\phi + \delta \phi)\) for the solution and contrast it with the lineal term. In the graph \(V_0\) is the greatest valor of $V$ over the interval $[0.01, 1]$. The initial conditions are \(\delta\phi(0.01) = 1\), \(\delta\phi'(0.01) = -1\), with \(\zeta = 0.25\), \(\phi_0 = 0\), \(\phi_1 = -1\) and \(\lambda = 10^{-3}\). The blue line is the solution potential and the red is the lineal approximation.}}\label{fig12}
\end{figure}

\section{Conclusions and Discussion}

The lack of gravitating effects on spacetime is the most impressive peculiarity of the stealth field. Moreover, there is no direct interaction between electromagnetic and stealth fields, which makes it harder to observe them. We wonder if there is the possibility of observing the stealth; we seek an answer about perturbing the stealth with the certainty that the perturbed field, in general, is not stealth either. In the present work--at difference with the orthodox perturbation where the spacetime point must perturbed too-- we demand that the background space and density remain invariant. The perturbation is the simplest one, \(\phi \to \phi + \lambda \delta\phi\), where we introduce an auxiliary scalar field \(\delta\phi\) and a parameter, \(0 \leq \lambda \leq 1 \), associated with the intensity of the perturbation. We pretend to determine whether these perturbations leave traces in spacetime and, if so, quantify them. 
%
%Additionally, due to the particular nature of the perturbed equation, we noted that the linear and quadratic coefficients in \eqref{14} appeared to be independent. Out of curiosity, we compared the linear coefficient with the total solution. We studied the four branches of solutions and showed in the graphs that for the range of interest, the numerical solutions of the linear term coincide quite well with the total solution. It is worth remembering that the scale factor \(a = 1\) represents the current value; for values \(0 < a < 1\), it corresponds to the time elapsed since the Big Bang until today. 
%

We solve equation (\ref{15}) using Runge-Kutta-Fehlberg (RKF45) methods to second-order ordinary differential equations \cite{Hairer2008} and show its solutions in Figures (1) to (4). However, analyzing the results, we note that these perturbations also work to discriminate between solutions. In branch 1, Fig.(1), the solution with $\lambda=1$ is closer to the exact, and in branch (4), it happens. However, branch 1 need more analysis to discard; but definitively, the branch 4 has an unphysical behavoir. 
Additionally, due to the particular nature of the perturbed equation (\ref{14}), we noted that it looks like a series expansion of the parameter, where the zero-order $\lambda$ term correspond to the original equation, the lineal term is a slight modification of the original equation, and curiously the quadratic term has the same equation as the zero-order. Because of this, we contrast the solutions between the linear term and the complete equation, and show the results in the Tables II-V. From these we observe that the contribution most important of solution coming from the linear term. Leaving the open question when and under what conditions it is possible to use this approximation.

\section*{Acknowledges}
The authors thank Dr. Giovany Cruz for the enlightening discussions and comments on this work. Special thanks are also extended to the SNI, the PRODEP program, CONAHCYT and CEFITEV-UV for their support in carrying out this work.

\newpage
%\appendix
%\section{Particular Stealth Field Solutions}
%The solutions of \ref{12} were presented in \ref{1}, the following Table provides a summary of these solutions.
%\begin{table}[htbp]
%\centering
%\small % Reduce font size 
%\renewcommand{\arraystretch}{1} % Increase row height
%\begin{tabular}{|c|c|c|}
%\hline
%\textbf{Range of $\zeta$} & \textbf{Solution} & \textbf{Constants and variables involved} \\ \hline
%$\zeta \in (-\infty, 0) \cup (12/73, \infty) - 1/4$ (Branch 1) 
%& $\phi(a) = \Phi_0 \left( \left(\frac{a}{a_0}\right)^{5/4+\beta} - \left( \frac{a}{a_0} \right)^{5/4-\beta} \right)^{\frac{2\zeta}{4\zeta -1}}$ 
%& $a_0 = \left( \frac{\phi_0}{\phi_1} \right)^{1/2\beta}$, \\
%& & $\beta = \frac{\sqrt{\zeta(73\zeta - 12)}}{4\zeta}$,\\ & & $\Phi_0 = \left(\phi_0 ^{\frac{5/4 + \beta}{2\beta}} -\phi_1 ^{\frac{5/4- \beta}{2\beta}}\right)^{\frac{2\zeta}{4\zeta -1}}$ \\ \hline
%
%$\zeta = 12/73$ (Branch 2) 
%& $\phi(a) = (\phi_1 \ln(a) - \phi_0)^{-24/25} a^{-6/5}$ 
%&  \\ \hline
%
%$0 < \zeta < 12/73$ (Branch 3) 
%& $\phi(a) = \phi_0 a^{5/4} \left( \phi_0 \sin(\gamma a) - \phi_2 \cos(\gamma a) \right)^{\frac{12\zeta}{73\zeta - 12}}$ 
%& $\gamma(a) =\frac{\sqrt{12-73\zeta}}{4\sqrt{3}}ln(a)$  \\ \hline
%
%$\zeta = 1/4$ (Branch 4) 
%& $\phi(a) = e^{-\phi a^{1/2}} a^{-3/2}$ 
%&  \\ \hline
%
%\end{tabular}
%\caption{The table presents the solutions from [Álvarez, Campuzano, Cárdenas 2019]. $\phi_0$ and $\phi_1$ are integration constants.}
%\label{table:solutions}
%\end{table}
%\vspace{-0.4cm}
%\begin{multicols}{2}

\end{multicols}
\end{document}